\documentclass[twocolumn,aps,prc,tightenlines,floats,floatfix,preprintnumbers,nofootinbib]{revtex4}

\def\sp{\kern +3pt}
\def\sm{\kern -3pt}
\def\spQ{\kern +6pt}

\def\bea{\begin{eqnarray}}
\def\eea{\end{eqnarray}}

\def\sfrac#1#2{{\textstyle \frac{#1}{#2}}}

\def\be{\begin{equation}}
\def\ee{\end{equation}}
\def\ba{\begin{eqnarray}}
\def\ea{\end{eqnarray}}

\usepackage{graphics}
\usepackage{graphicx}
\usepackage{epsf}
\usepackage{amsmath}
\usepackage{amssymb}

\begin{document}

\phantom{0}
\vspace{-0.2in}
\hspace{5.5in}

\preprint{}

\vspace{-1in}

\title
{\bf A covariant model for the $\gamma^\ast N \to  N^\ast(1520)$ reaction}
\author{G.~Ramalho$^1$ and
M.~T.~Pe\~na$^{2}$ \vspace{-0.1in}}

\affiliation{
$^1$International Institute of Physics, Federal 
University of Rio Grande do Norte, Av.~Odilon Gomes de Lima 1722, 
Capim Macio, Natal-RN 59078-400, Brazil
\vspace{-0.15in}}
\affiliation{
$^2$CFTP, Instituto Superior T\'ecnico, Universidade de Lisboa, \\
Av.~Rovisco Pais, 1049-001 Lisboa, Portugal}

\vspace{0.2in}
\date{\today}

\phantom{0}

\begin{abstract}
We apply the covariant spectator quark model 
to the study of the electromagnetic structure of the $N^\ast(1520)$ 
state ($J^{P}= \frac{3}{2}^-$),
an important resonance from the 
second resonance region in both spacelike and timelike regimes.
The contributions from the valence quark effects 
are calculated for the $\gamma^\ast N \to N^\ast(1520)$ 
helicity amplitudes.
The results are used to 
parametrize the meson cloud dominant at low $Q^2$.
\end{abstract}

\vspace*{0.9in}  
\maketitle

The electromagnetic structure of the 
nucleon ($N$) and the nucleon excitations ($N^\ast$) can be probed 
through the $\gamma^\ast N \to N^\ast$ reactions, 
with squared momentum transfer $q^2 < 0$ 
(spacelike region).
Experimental facilities such as Thomas Jefferson Laboratory (Jlab),
MIT-Bates and Mainz provide nowadays important 
information about those reactions for 
low and high $Q^2$, where $Q^2= - q^2 > 0$~\cite{Aznauryan12a,NSTAR,Burkert04}.

In order to interpret the data one has 
to rely on theoretical models 
based either on the fundamental 
QCD degrees of freedom, quarks and gluons,
or effective ones.
Although the microscopic dynamics 
refers to quarks and gluons, those degrees of freedom can be  observed 
only at very high $Q^2$.
At low and intermediate $Q^2$ 
more phenomenological descriptions 
with baryon and mesons 
or constituent quarks  
can be justified~\cite{Aznauryan12a,Burkert04,Capstick00,NSTAR}.

Among the constituent quark models the covariant 
spectator quark model (CSQM) \cite{NSTAR,ExclusiveR,Nucleon,Omega,Octet}
was successfully applied 
to the nucleon \cite{Nucleon}, the $\Delta(1232)$ 
\cite{Delta},
to several others nucleon resonances such as 
the $N^\ast (1440)$ \cite{Roper}, the $N^\ast(1535)$ \cite{S11}
and other baryons  \cite{DecupletDecays,Others}.
We apply now the CSQM to the $N^\ast(1520)$ state 
and $\gamma^\ast N \to N^\ast(1520)$ transition.

The state  $N^\ast(1520)$ ($J^{P}= \sfrac{3}{2}^-$)
is an important state,
as it is the $N^\ast(1535)$ ($J^{P}= \sfrac{1}{2}^-$),
from the second resonance region,
and plays also an important role in the timelike region ($Q^2 < 0$).
The extension of the CSQM to the timelike region was 
already made for the  $\Delta(1232)$ \cite{Timelike}.
The available data 
for the $\gamma^\ast N \to N^\ast(1520)$ transition, 
combined with model estimations 
suggest that valence quark effects
dominate at high $Q^2$, 
while effects of the meson cloud dressing of the 
baryons can be significant at low $Q^2$~\cite{Aznauryan12a,NSTAR,CLAS,Diaz08}.

In the CSQM the wave function of the baryon $B$, 
$\Psi_B$,  is determined 
by the baryon properties 
(flavor, spin, orbital angular momentum, etc.),
and their symmetries, and the radial part 
represented by a scalar function $\psi_B$
adjusted phenomenologically to 
the experimental electromagnetic form factor data, 
and lattice QCD data for some  baryon systems~\cite{ExclusiveR,Octet,Omega}.
According to the spectator theory, two of the 
quarks can be considered on-shell in the intermediate 
states and the third quark is free to interact 
with the electromagnetic probe.
Integrating in the degrees of freedom of the 
quark-pair we can reduce the baryon to 
a quark-diquark system where the diquark
is on-shell with effective mass $m_D$ 
\cite{Nucleon,Omega,Nucleon2}.
The quark electromagnetic current is described 
using vector meson dominance (VMD).
The quark electromagnetic form factors
are parametrized by some vector meson masses~\cite{Nucleon,Octet,Omega}.
The vectors meson poles parametrize 
the {\it spatial extension} of the constituent quarks. 
VMD is very useful for 
the extrapolation of the model to other regimes 
such as the lattice regime \cite{Lattice} and 
the timelike regime~\cite{Timelike}.
Finally the electromagnetic current 
between two baryon states is obtained 
by taking the impulse approximation 
and summing in the individual 
quark currents~\cite{ExclusiveR,Nucleon,Nucleon2,Omega}.

For the transition between the nucleon and 
the $N^\ast (1520)$ state, here labeled as $R$
(for resonance), we need to construct  
the nucleon and the $R$ wave functions.
For the nucleon, we use the wave function 
derived in a work  
where the nucleon is described as a $S$-wave 
quark-diquark system \cite{Nucleon}.
For the $R$ wave function we took  
a covariant generalization 
of the non relativistic form \cite{Capstick00,D13}.
The  non relativistic form 
couples angular momentum states 
$Y_{1m}$ ($m=0, \pm 1$) 
in the diquark momentum $k$ and 
in the relative quark-pair momentum $r$ 
to  three-spin states with different symmetries 
(mixed-symmetric, mixed-antisymmetric and totally symmetric).
In principle the $N^\ast (1520)$ wave function
is a combination of states with core spin
(sum of the quark spins) 1/2 and 3/2,
but hadronic decays suggests that the spin-3/2
admixture is small ($\sin \theta_D \approx 0.1 \ll 1$) 
\cite{Capstick00}.

In the transition between a $J^P= \frac{1}{2}^+$ (nucleon) 
and a $J^P= \frac{3}{2}^-$, like the $N^\ast(1520)$ state, one 
defines three independent electromagnetic form factors.  
In particular one 
can consider $G_M$, the magnetic dipole form factor, 
the function $G_4^\prime = -(G_M + G_E)$,
where $G_E$ is the electric quadrupole form factor,
and $G_C$, the Coulomb quadrupole form factor.
Alternatively, one can use 
the {\it classic} helicity amplitudes, $A_{1/2}$, $A_{3/2}$ 
and $S_{1/2}$, in the $R$ rest frame.
One has then~\cite{D13}
\ba
A_{1/2} \propto G_M + \frac{1}{4}G_4^\prime, 
\hspace{.4cm} 
 A_{3/2} \propto G_4^\prime ,
\hspace{.4cm} 
S_{1/2} \propto G_C. 
\label{eqAmps}
\ea
The form factors and the helicity amplitudes 
are functions of $Q^2$ only.

In the CSQM the transition form factors
are written as a combination 
of quark electromagnetic form factors 
and the radial wave functions.
The dependence on the radial 
wave functions is  
codified in a covariant function  $I_z(Q^2)$,
which is an {\it overlap integral}.
In the $R$ rest frame one has
\ba
I_z(Q^2)=
\int_k \frac{k_z}{|{\bf k}|}
\psi_R(P_R,k) \psi_N(P_N,k) 
\label{eqInt},
\ea 
where $P_R$ ($P_N$) are the $R$ (nucleon) momentum 
and $k$ the diquark momentum at the $R$ rest frame.
The integration symbol is a short notation 
for the covariant integration in $k$.
A covariant generalization of the equation (\ref{eqInt})
can be derived in an arbitrary frame \cite{D13}. 

\begin{figure}
\centering
\begin{centering}
\includegraphics[width=6cm,clip]{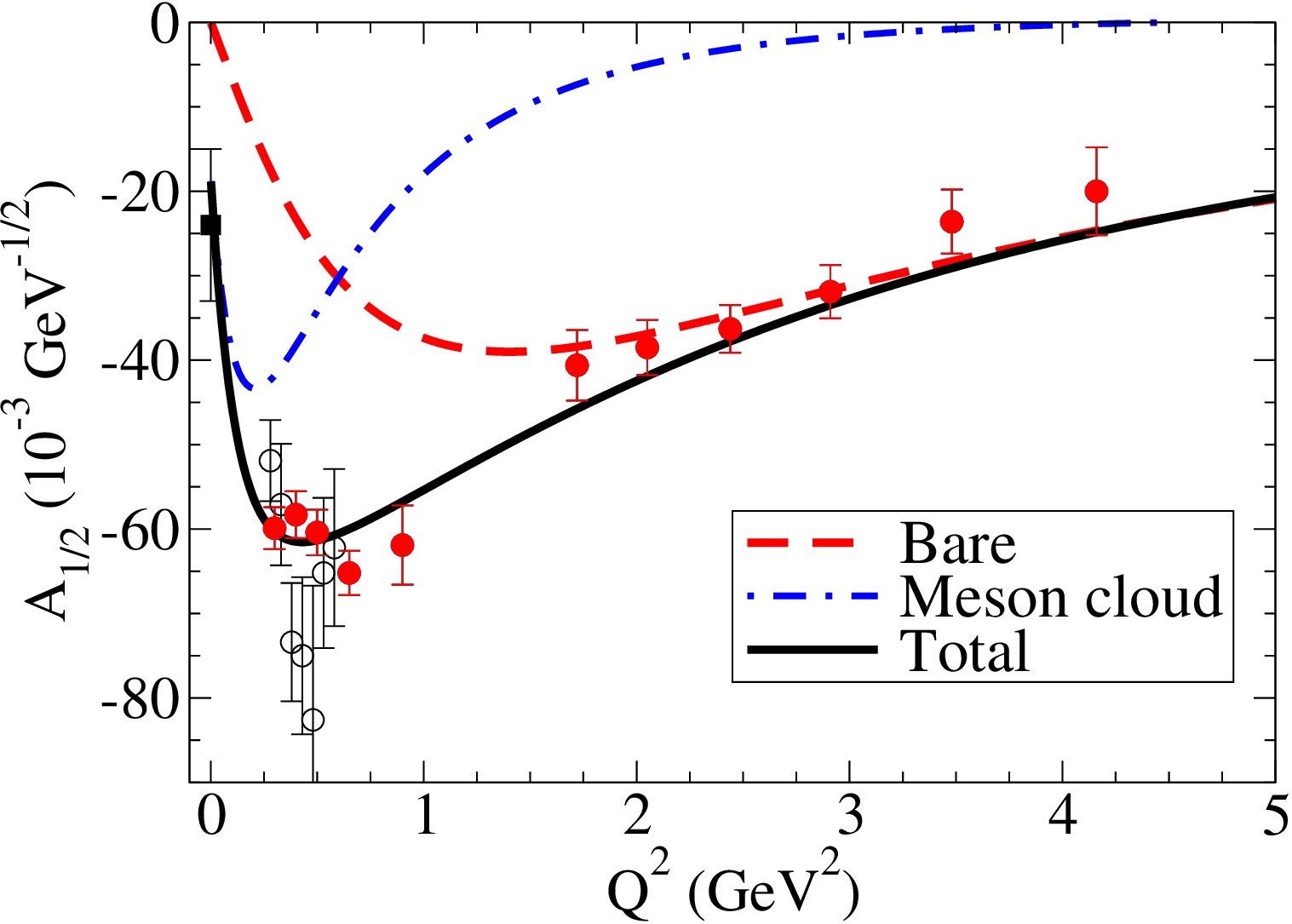}
\includegraphics[width=6cm,clip]{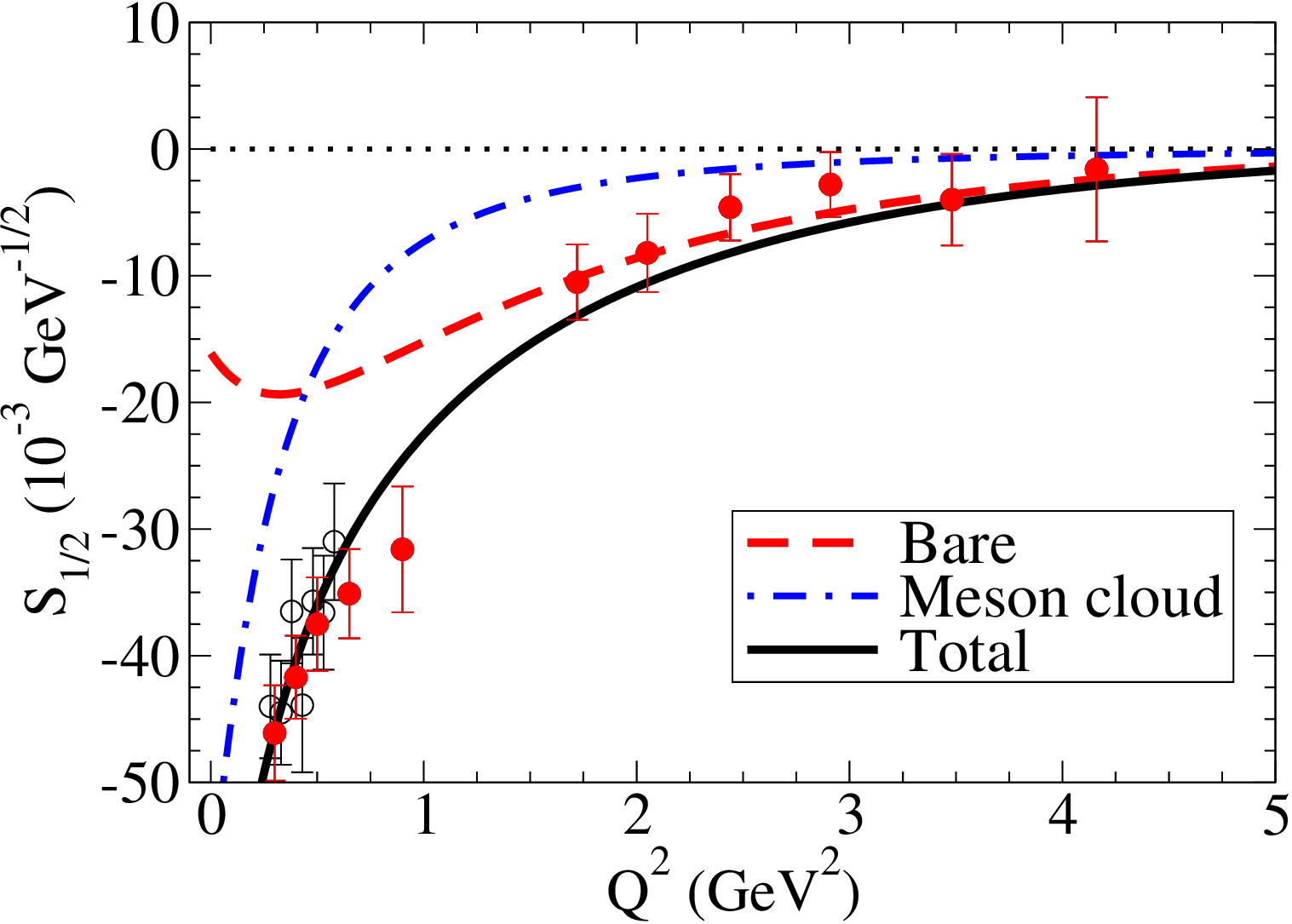} 
\end{centering} \\
\begin{centering}
\includegraphics[width=6cm,clip]{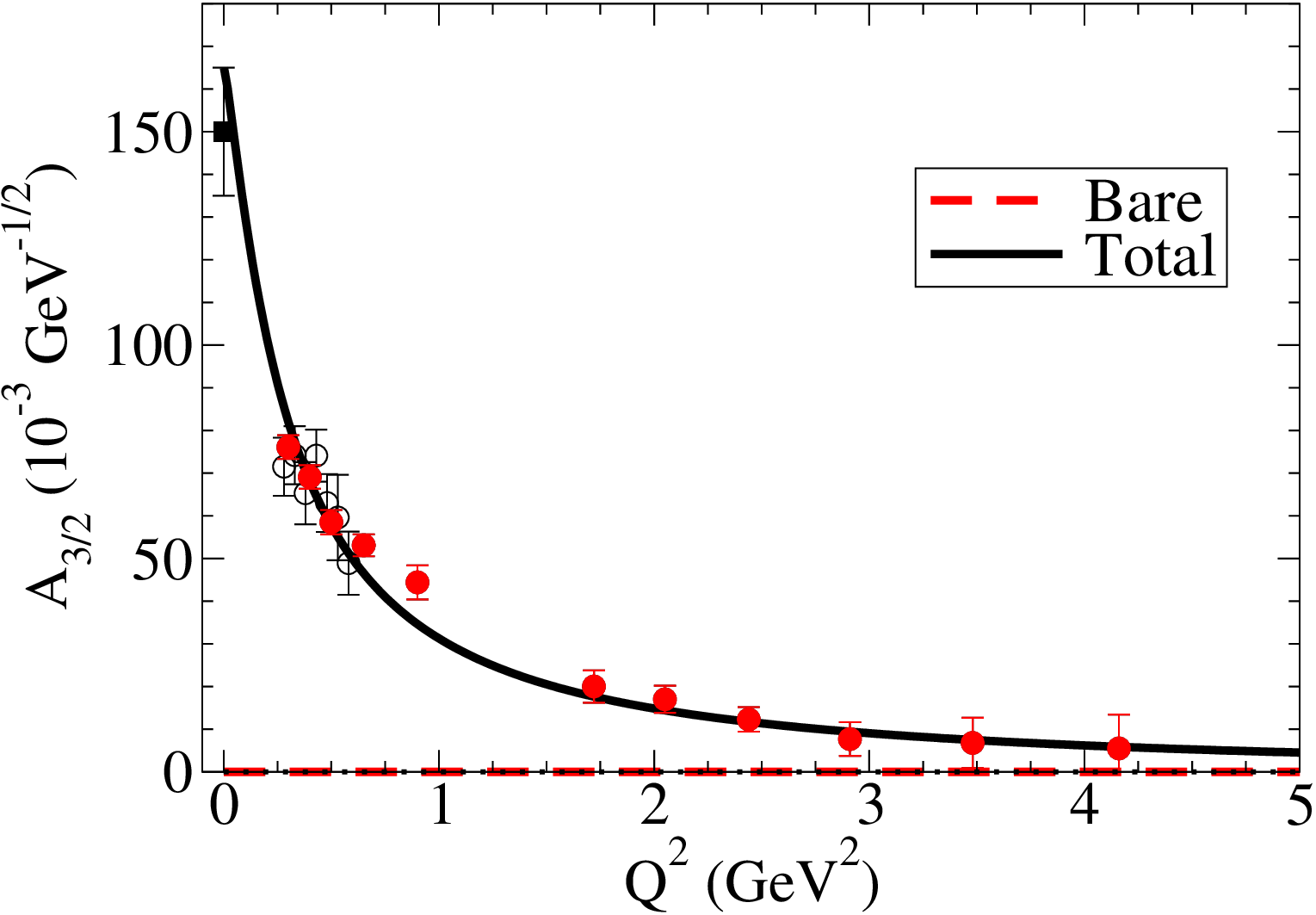}
\end{centering}
\caption{Helicity amplitudes for the 
$\gamma^\ast N \to N^\ast(1520)$ at the resonance rest frame.
Valence quark (dash-line), meson cloud effects (dot-dashed line)
and total (solid line).
Data from CLAS/Jlab \cite{CLAS}. 
Not included are the results from the MAID analysis
\cite{Tiator09}.
}
\label{fig1}       
\end{figure}

The  orthogonality between the nucleon and the 
$R$ state is expressed by the condition $I_z(0) =0$.
Based on this condition it is possible to 
construct a function $\psi_R(P_R,k)$  
with one adjustable parameter that 
can be fitted to the data.
The calculations using the spin-1/2 
component of the $R$ wave function lead to
\ba
A_{1/2} \propto I_z(Q^2), 
\hspace{.4cm} 
 A_{3/2} \equiv 0,
\hspace{.4cm} 
S_{1/2} \propto \frac{I_z(Q^2)}{Q^2}. 
\ea
According to equation~(\ref{eqAmps}) these results are a consequence 
of $G_4^\prime \equiv 0$.
The results for the valence quark contributions to the  
helicity amplitudes
are presented in the figure~\ref{fig1} (dash-lines).
Our fit of  the function $\psi_R$ was made for $Q^2 > 1.5$ GeV$^2$, 
a regime where we expect small meson cloud effects.

\begin{figure}
\includegraphics[width=6cm,clip]{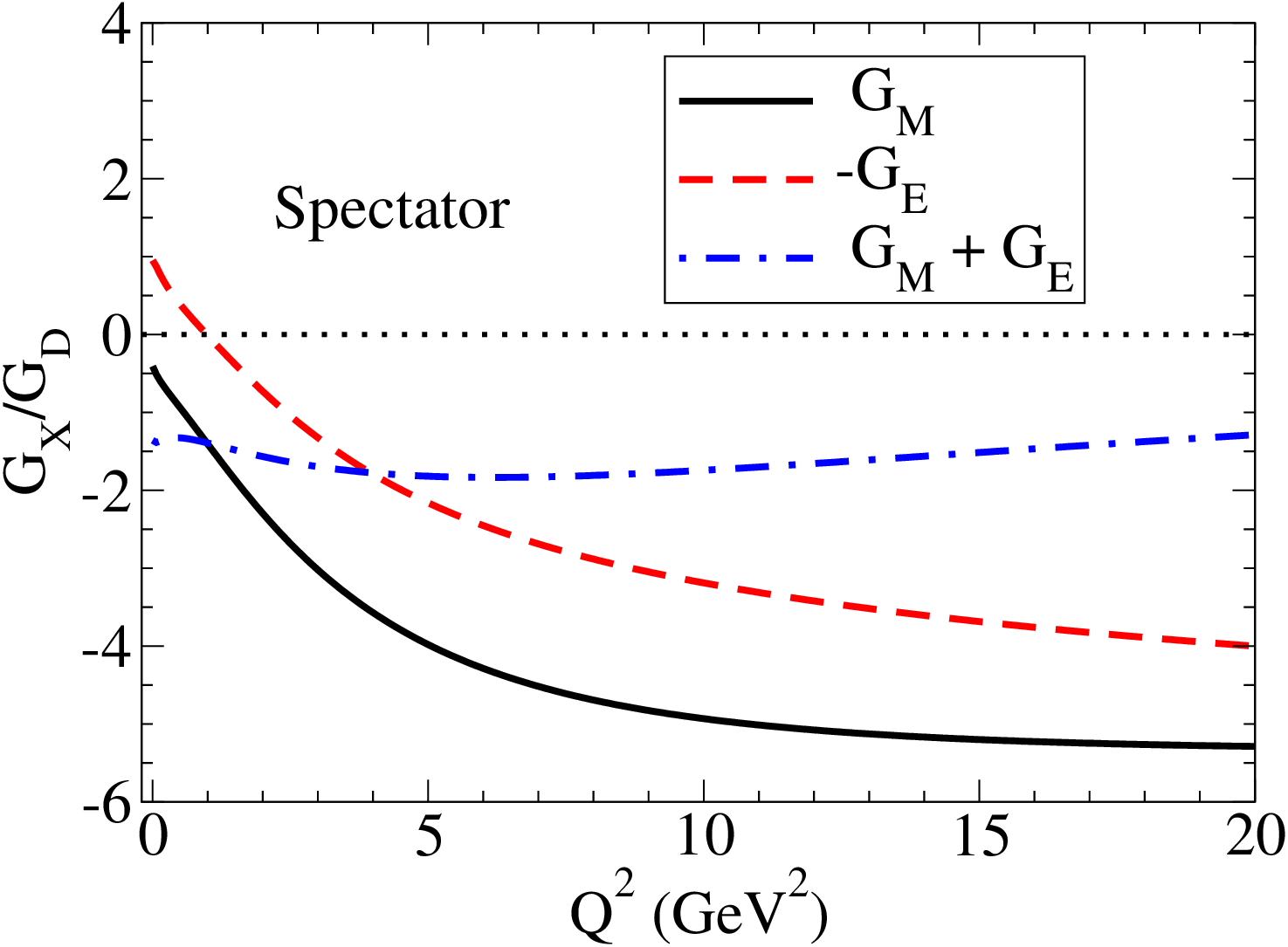}
\caption{Results of the form factors $G_M$, $-G_E$ 
for very high $Q^2$. 
Scale extended to 20 GeV$^2$ 
(prediction to the Jlab-20 GeV update regime).
Form factors normalized with $G_D= \left( 1 + \frac{Q^2}{0.71} \right)^{-2}$,
where $Q^2$ is in GeV$^2$. 
Note the slow falloff of $G_M + G_E $ with $Q^2$.
}
\label{fig2}       
\end{figure}

Although in the CSQM the quarks have 
structure, there are processes such as 
meson exchange between two different 
quarks inside the baryon, 
that cannot be interpreted just 
as quark dressing and have to be classified
at hadronic level as meson cloud~\cite{DecupletDecays,Octet}.
Assuming then that the CSQM gives  
a good description of the valence quark effects only, 
we used the model to extract the meson cloud 
contributions from the data.
Since the pion is the dominant decay,
we assumed a parametrization regulated 
by the pion 
with some effective momentum ranges (cutoffs)
adjusted to the global data.
The results of the fit are presented also 
in the figure~\ref{fig1} (dot-dashed lines).
The solid line gives the final result (total).

From figure~\ref{fig1}, one concludes that 
valence quark effects dominate the high $Q^2$ regime 
($Q^2 > 1.5$ GeV$^2$) for the amplitudes $A_{1/2},S_{1/2}$, 
although meson cloud effects are significant at low $Q^2$.
As for $A_{3/2}$, only the meson cloud contributes.
This result differs from other quark model estimates, 
but it is consistent with the large meson cloud 
estimate from the EBAC group~\cite{Diaz08}.

\newpage

Finally we look at the results for the 
multipole form factors at very high $Q^2$,
extending the model to the region 
of the Jlab 12-GeV upgrade. 
In the figure~\ref{fig2}, we plot $G_M$, $-G_E$ and 
$G_M+ G_E,$ normalized by the dipole form factor $G_D$,
for a  better observation of the falloff of those form factors.
Note the scaling of  $G_M$, $-G_E$,
and also the very slow falloff of $G_M + G_E$,
that is negligible for very high $Q^2$, according 
to pQCD \cite{Carlson} and also to the CSQM.
Those results suggest that $G_M + G_E \simeq 0$ 
(or $G_E \simeq -G_M$), equivalent to $A_{3/2} \simeq 0$,
happens only for very large $Q^2$.

\newpage

The present parametrization of the 
$\gamma^\ast N \to N^\ast(1520)$ transition form factors 
in the spacelike regime can be extended to 
the timelike regime \cite{NewPaper}.

\vspace{.1cm}

{\bf Acknowledgments:}
This work was supported by the Brazilian Ministry of Science,
Technology and Innovation (MCTI-Brazil), and
Conselho Nacional de Desenvolvimento Cient{\'i}fico e Tecnol\'ogico
(CNPq), project 550026/2011-8.
MT~Pe\~na received financial support from Funda\c c\~ao 
para a Ci\^encia e a Tecnologia (FCT) under Grants 
Nos.~PTDC/FIS/113940/2009, CFTP-FCT 
(PEst-OE/FIS/U/0777/2013) and POCTI/ISFL/2/275. 
This work was also partially supported by the European Union 
under the HadronPhysics3 Grant No.~283286.

\end{document}